\documentclass[runningheads]{llncs}
\usepackage[T1]{fontenc}
\usepackage{graphicx}
\begin{document}
\title{AICoFe Demo: AI-based Collaborative Feedback System for Peer, Self, and Teacher Assessment}
\titlerunning{AICoFe Demo: AI-based Collaborative Feedback System}
%
\author{Alvaro Becerra\inst{1}\orcidID{0009-0003-7793-2682} \and
Alejandra Palma\inst{1}\orcidID{0009-0005-6555-7250} \and
Ruth Cobos\inst{1}\orcidID{0000-0002-3411-3009} \and Julian Fierrez\inst{2}\orcidID{0000-0002-6343-5656}}
\authorrunning{A. Becerra et al.}
\institute{GHIA, School of Engineering, Universidad Autónoma de Madrid, Spain \and BiometricsAI, School of Engineering, Universidad Autónoma de Madrid, Spain}

\maketitle              
\begin{abstract}
Peer feedback promotes active learning, critical reflection, and skill development, but its effectiveness is often limited by the quality of feedback students provide. Recent advances in LLMs offer new opportunities to support peer feedback by generating more coherent and actionable feedback while preserving human oversight. This paper presents AICoFe, an AI-based collaborative feedback system designed to support teacher, peer, and self-assessment in higher education. AICoFe integrates rubric-based evaluations, GenAI-supported feedback, Learning Analytics dashboards, and video recordings to foster reflective learning. The system combines quantitative scores and qualitative observations to generate structured feedback focused on strengths, areas for improvement, and actionable recommendations, which teachers can review and curate. An evaluation with 65 undergraduate and master’s students shows high satisfaction with the coherence and usefulness of the feedback, as well as the excellent usability, indicating that AICoFe effectively supports peer feedback in authentic educational settings.

\end{abstract}
\section{Introduction}
Peer feedback promotes active learning, critical reflection, and a deeper understanding of quality criteria~\cite{topping2009peer,liu2006peer}. By acting as evaluators, students strengthen their self-regulation and evaluative judgment~\cite{van2017exploring}. 
Despite its educational potential, peer feedback is often superficial or inconsistent, as students often lack the feedback literacy needed to produce constructive comments~\cite{wei2024incorporating,viberg2024exploring}.

Recent advances in Generative Artificial Intelligence (GenAI) open new opportunities to address these challenges and enhance peer feedback processes. Previous approaches to automatic feedback relied mainly on predefined rules, templates, or numerical indicators \cite{cavalcanti2021automatic,bodily2018design}. These approaches often produced generic, decontextualized feedback that was difficult for students to interpret or act on, limiting their impact on learning. More recently, large language models have demonstrated the ability to generate coherent, context-aware, and personalized feedback based on both quantitative data and qualitative input \cite{giannakos2024promise}. In educational settings, GenAI has been successfully applied to improve the clarity, usefulness and actionability of feedback, helping students understand their performance and identify concrete steps for improvement \cite{zhang2024students,becerra2024generative}.

Building on this perspective, this paper presents an improved version of AICoFe ``Artificial Intelligence-based Collaborative Feedback'' system \cite{becerra2025enhancing}. AICoFe is a human-centered system that supports human–AI collaboration through role-specific dashboards \cite{becerra2025m2lads} and teacher-in-the-loop feedback mediation. The system enables both teachers and students to conduct evaluations using customized rubrics that integrate quantitative scores and qualitative observations. Powered by generative AI (GenAI), AICoFe generates personalized feedback for each student, providing actionable insights to improve their skills. The Learning Analytics (LA) dashboards of the system offer dynamic features, including personalized summaries, comparative graphs, and self-assessment tools. In addition, video recordings of student performances support reflective learning.\footnote{Demonstration of the entire AICoFe workflow: \url{https://youtu.be/wPggcyz7Olk}}

\section{AICoFe: AI-based Collaborative Feedback System}

AICoFe is organized into four main modules: (i) the Visualization Module (see Sect.~\ref{ss:visualization_module}), (ii) the Management Module, (iii) the Feedback Generation Module (see Sect.~\ref{ss:feedback_module}), and (iv) the Recording Module. The Management Module provides connectivity between the platform and its data storage infrastructure, including SQL and MongoDB databases as well as the file system. The Recording Module supports reflective learning by capturing students' performances in video and audio format during evaluation activities, using a standard laptop with Microsoft Teams as the recording platform. 

\subsection{Visualization Module}\label{ss:visualization_module}

AICoFe provides a set of interactive Learning Analytics dashboards developed with the Dash framework to support the different actors involved in the peer feedback workflow.

\begin{figure}[t]
  \centering
  \begin{minipage}{0.49\linewidth}
    \centering
    \includegraphics[height=5cm]{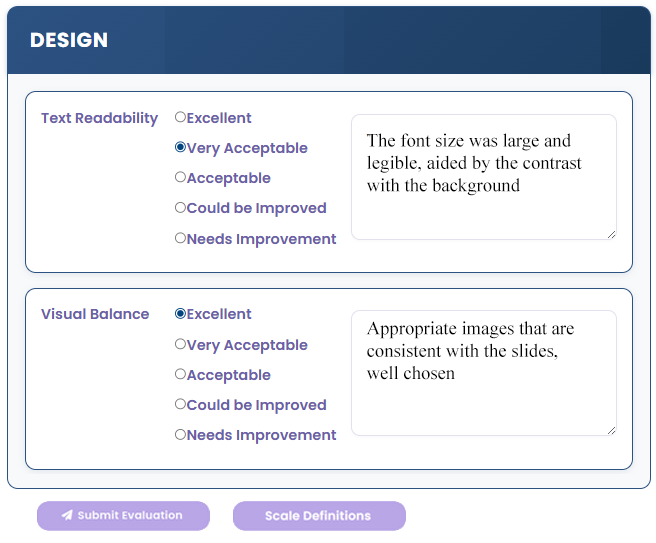}
    \par\small (a)
  \end{minipage}
  \hfill
  \begin{minipage}{0.49\linewidth}
    \centering
    \includegraphics[height=5cm]{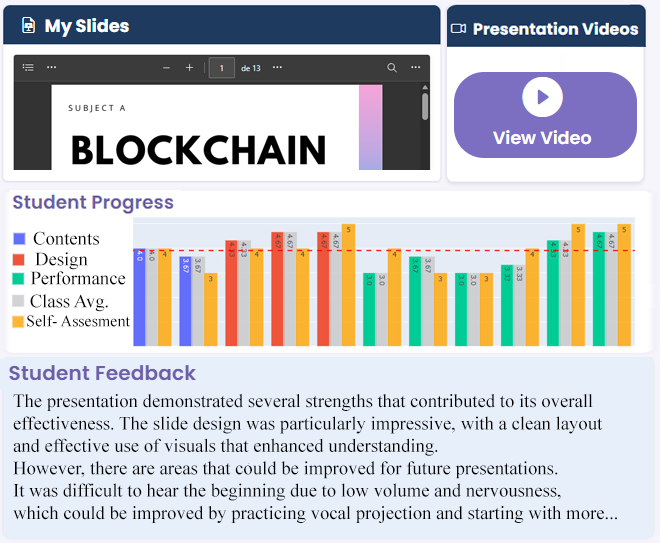}
    \par\small (b)
  \end{minipage}
  \caption{AICoFe dashboards: (a) Evaluator Dashboard, (b) Student Dashboard.}
  \label{fig:ev_student}
\end{figure}

The \textbf{Evaluator Dashboard} (see Fig.~\ref{fig:ev_student} (a)) supports teachers and students during evaluation. Evaluators select a rubric and a target student, assign Likert scores per item, and add written comments, with rubric level descriptions 
always visible to align interpretations. The system also captures interaction metadata, such as score timing, for later analysis.

The \textbf{Student Dashboard} (see Fig.~\ref{fig:ev_student} (b)) is designed to promote reflective learning through self-assessment and feedback exploration. Students can revisit a video recording of their performance and complete a self-evaluation using the same rubric applied by peers and instructors. Once completed, the dashboard visualizes the differences and alignments between the self-assessment scores and the aggregated evaluations received. Alongside these visual elements, students receive AI-supported collaborative feedback. Students can also rate the coherence and usefulness of the feedback directly within the dashboard.

\begin{figure}[t]
    \centering
    \includegraphics[width=1\linewidth]{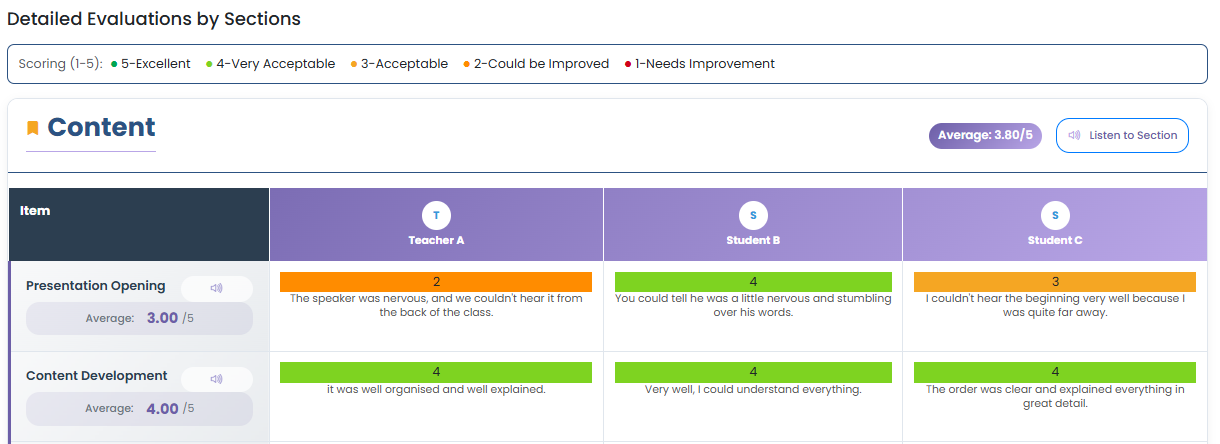}
\caption{Teacher Dashboard with quantitative scores and qualitative observations.}

    \label{fig:detailed_evaluation}
\end{figure}

\begin{figure}[t]
    \centering
    \includegraphics[width=1\linewidth]{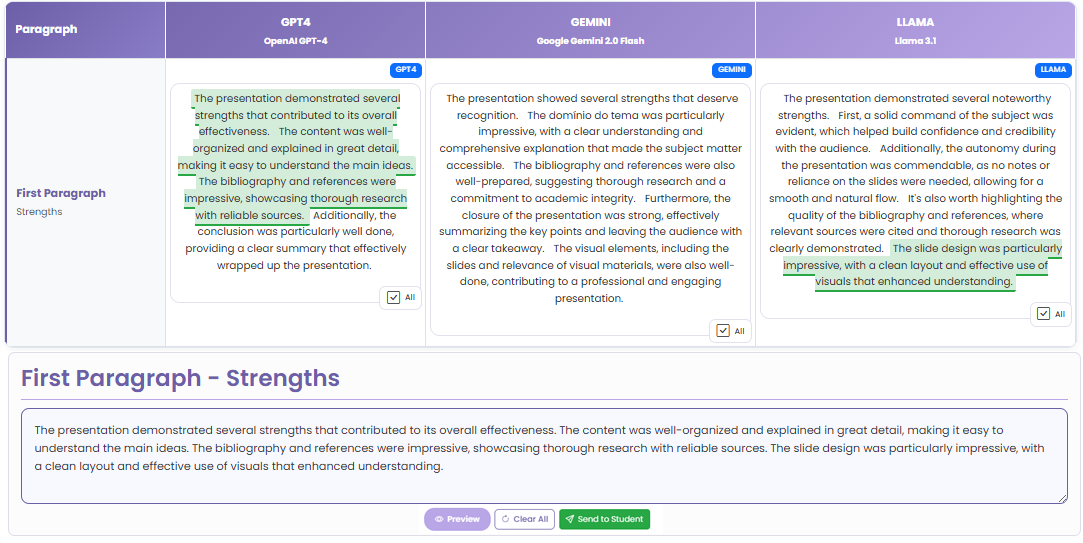}
\caption{Screenshot of the Teacher Dashboard showing feedback generated by different LLMs and teacher-mediated composition of final feedback.}

    \label{fig:granular_feedback}
\end{figure}

The \textbf{Teacher Dashboard} (see Fig.~\ref{fig:detailed_evaluation}) supports feedback review and curation. Teachers can inspect quantitative scores and qualitative comments from peers and instructors, access student performance videos, and review feedback generated by the different LLMs (see Fig.~\ref{fig:granular_feedback}). Teachers can select, edit, or combine sentences from AI-generated outputs to compose the final feedback. A log tracks drafts and sent versions, distinguishing LLM sentences from teacher-modified sentences. The dashboard also integrates text-to-speech (Google Text-to-Speech) that enables teachers to listen to detailed evaluation comments (see Fig.~\ref{fig:detailed_evaluation}).

The system configuration and evaluation orchestration are handled through an \textbf{Admin Dashboard}.

\subsection{Feedback Generation Module}\label{ss:feedback_module}
The Feedback Generation Module is responsible for producing personalized feedback by combining quantitative assessment results from teachers and peers with qualitative observations. The module builds upon previous work \cite{becerra2024generative} and extends it to a multi-model setting in which three independent feedback instances are generated, one per LLM. Each feedback instance follows a predefined structure composed of three paragraphs: (i) strengths, highlighting well-performed aspects; (ii) areas for improvement, identifying specific weaknesses; and (iii) an action plan, providing concrete recommendations for future improvement

As an initial step, the module preprocesses the qualitative input collected from the evaluators. Textual comments are normalized to remove invalid characters and are automatically screened to assess their relevance with respect to the evaluated rubric item. Filtered data are combined into course-specific prompts that integrate average rubric scores and validated qualitative observations. Prompts also incorporate rubric level descriptions. All data sent to the generative models are anonymized. The module supports the use of 3 large LLMs: OpenAI GPT-4.1-mini, Google Gemini 2.5 Flash, and Meta Llama 3.1. 

All three models are fine-tuned using supervised learning on the SOPHIAS dataset~\cite{becerra2026sophias}, on input--output pairs structured according to the three-paragraph feedback schema. GPT-4.1-mini was fine-tuned via the OpenAI API for 3 epochs (lr multiplier: 2, final loss: 0.934). Gemini 2.5 Flash was fine-tuned via Google Vertex AI for 40 cycles, with the optimal checkpoint selected at step 45 (final loss: 0.759, accuracy: 0.744). Llama 3.1 was fine-tuned locally via LoRA ($r=16$, $\alpha=32$) with 4-bit NF4 quantization for 3 epochs (final loss: 1.210).

A post-generation validation stage checks teacher-defined constraints (length, coherence, restricted terms), triggering automatic regeneration if needed. Regarding factual grounding, model prompts contain exclusively anonymized rubric scores and evaluator comments. Iterative prompt refinement ensures that the output remains grounded in the observations, mitigating hallucination risks.

\section{AICoFe Evaluation}
AICoFe was deployed during 2024--2025 with 46 undergraduate and 19 master's students at the UAM, who delivered oral presentations evaluated through peer, self, and teacher 
assessments using rubric-based criteria. The instructor design the rubrics based on the specific competency criteria of each subject, ensuring alignment with the learning objectives of each course. The reliability of this multi-source design was validated through inter-rater agreement analyses and scoring bias checks~\cite{becerra2025leveraging}, consistent with previous work~\cite{falchikov2000student,tang2023comparison,ross2006reliability}.

To evaluate student satisfaction, participants rated \textit{feedback coherence} and \textit{feedback usefulness} using numerical scales within the Student Dashboard. Master's students (17/19, 89\%) reported mean coherence of 9.24 ($\sigma = 0.83$) and usefulness of 9.00 ($\sigma = 1.22$). Undergraduate 
students (46/46, 100\%) reported a mean coherence of 8.83 ($\sigma = 0.92$) and a usefulness of 9.20 ($\sigma = 1.02$). High scores in both groups suggest that teacher-curated AI feedback was perceived as aligned with student performance and actionable.

Usability was measured using the System Usability Scale (SUS). A total of 31 participants completed the questionnaire, resulting in a mean SUS score of 
79.91 ($\sigma = 11.92$), which corresponds to an excellent usability level.

\section{Conclusions and Future Work}

This paper presented AICoFe, an AI-based collaborative feedback system for peer, self, and teacher assessment in higher education. By integrating rubric-based evaluation, AI-supported feedback, and Learning Analytics dashboards, AICoFe supports coherent, useful, and pedagogically meaningful feedback while preserving strong teacher oversight.

The results show high student satisfaction with the coherence and usefulness of the feedback across undergraduate and master’s levels, as well as an excellent SUS usability score. These findings highlight the importance of the teacher’s role as an active mediator in AI-assisted assessment. AICoFe also offers a practical and research-informed model to scaling peer assessment without compromising feedback quality. Future work includes a larger deployment and the integration of multimodal data and biometrics in the feedback process~\cite{becerra2025mosaicf,becerra2025review,becerra25multimobile,daza25multimobile}

\subsubsection{Acknowledgements.}

Support by Cátedra ENIA UAM-VERIDAS en IA Responsable (NextGenerationEU PRTR TSI-100927-2023-2), M2RAI (PID2024-160053OB-I00, MICIU/FEDER), Emotional (UAM AVANZA 2026) and SNOLA (RED2022-134284-T). A. Becerra is funded by an FPI from the Comunidad de Madrid (PIPF-2024/COM-34288).

\bibliographystyle{splncs04}
\bibliography{sample-base}

\end{document}